\documentclass[sigconf, authorversion=true]{acmart}
\usepackage[english]{babel}
\usepackage[utf8]{inputenc}
\usepackage[T1]{fontenc}
\usepackage{setspace}
\usepackage{tabulary} 
\usepackage{dblfloatfix}
\usepackage{framed} 
\usepackage{xcolor} 

\usepackage{csvsimple}

\usepackage{graphicx}

\PassOptionsToPackage{hyphens}{url}
\usepackage{hyperref}
\hypersetup{
    pdfauthor={Sascha El-Sharkawy, Adam Krafczyk and Klaus Schmid},
    pdftitle={MetricHaven - More Than 23,000 Metrics for Measuring Quality Attributes of Software Product Lines},
    pdfsubject={23rd International Conference on Software Product Line},
    pdfkeywords={Software Product Lines; SPL; Metrics; Implementation; Variability Models; Feature Models},
    pdfpagemode={UseOutlines},
    pdfproducer={LaTeX},
    colorlinks=true,
    linkcolor=black,          
    citecolor=black,          
    filecolor=black,          
    urlcolor=black            
}

\expandafter\def\expandafter\UrlBreaks\expandafter{\UrlBreaks
  \do\a\do\b\do\c\do\d\do\e\do\f\do\g\do\h\do\i\do\j%
  \do\k\do\l\do\m\do\n\do\o\do\p\do\q\do\r\do\s\do\t%
  \do\u\do\v\do\w\do\x\do\y\do\z\do\A\do\B\do\C\do\D%
  \do\E\do\F\do\G\do\H\do\I\do\J\do\K\do\L\do\M\do\N%
  \do\O\do\P\do\Q\do\R\do\S\do\T\do\U\do\V\do\W\do\X%
  \do\Y\do\Z}

\usepackage{enumitem}
\setlist[itemize]{leftmargin=1em}
\setlist[enumerate]{leftmargin=3ex}
\setlist[description]{leftmargin=1em}

\colorlet{shadecolor}{gray!25}
\newcommand{\Command}[1]{
    \begin{snugshade*}
      \noindent #1
    \end{snugshade*}
}

\newcommand{\trim}{\looseness-1}

\newcommand{\ifDef}{\texttt{\#ifdef}}

\newcommand{\ifDefBlock}{\ifDef-block}

\newcommand\footnoteref[1]{\protected@xdef\@thefnmark{\ref{#1}}\@footnotemark}

\newcommand{\SD}{SD}
\newcommand{\SDvp}{\SD$_\text{VP}$}
\newcommand{\SDfile}{\SD$_\text{File}$}

\newcommand{\NoF}{NoC}
\newcommand{\NoFOut}{\NoF$_\text{out}$}
\newcommand{\NoFIn}{\NoF$_\text{in}$}

\newcommand{\CoC}{CoC}

\newcommand{\tRef}[1]{\tikz[remember picture]\node[](#1){};\hspace{-5pt}}
\usepackage{tikz}
\usetikzlibrary{matrix,arrows,positioning,shadows}
\usetikzlibrary{shapes.arrows,fit,shapes.geometric}
\usetikzlibrary{chains}
\tikzset{%
  block/.style = {draw=red!70!black, thick, rectangle},
}

\definecolor{cEclipseRed}{rgb}{0.6,0,0}
\definecolor{cEclipseGreen}{rgb}{0.25,0.5,0.35}
\definecolor{cEclipsePurple}{rgb}{0.5,0,0.35}
\definecolor{cEclipseBlue}{rgb}{0.25,0.35,0.75}

\usepackage{listings}
\lstdefinelanguage{config} {
  language=C,
  morecomment=[l]{\#},
  morekeywords=[2]{y, n, m},
  keywordstyle=[2]{\color{red!70!black}}
}

\lstdefinelanguage{kconfig} {
  language=config,
  string=[b]",
  keywords = {choice, endchoice, config, bool, tristate, string, int, hex, select, depends, on, option, modules, default, def_bool, def_tristate, range, if, endif, menu, endmenu, prompt, source, help}
}

\lstdefinelanguage{kbuild} {
  language=[gnu]make,
	moredelim=[is][stringstyle]{@<}{>@}
}

\lstdefinelanguage{cWithPre} {
  language=C,
  morecomment=[l][\bfseries\color{blue!60!black}]{\#},
}

\lstdefinelanguage{cWithPre2} {
  language=cWithPre,
  numbers=none,
	xleftmargin=.25em,
}

\lstdefinelanguage{pseudo} {
    language=Java,
    morekeywords=[2]{end, done, begin},
    mathescape
}

\lstdefinelanguage{properties} {
    tabsize=1,
    commentstyle=\color{blue}, 
    morecomment=[s][\color{cEclipsePurple}]{<}{>},
		morecomment=[l]{\	},
    morecomment=[l][\color{cEclipseGreen}]{\#},		
}

\setcopyright{acmlicensed}
\copyrightyear{2019} 
\acmYear{2019} 
\acmConference[SPLC '19]{23rd International Systems and Software Product Line Conference - Volume B}{September 9--13, 2019}{Paris, France}
\acmBooktitle{23rd International Systems and Software Product Line Conference - Volume B (SPLC '19), September 9--13, 2019, Paris, France}
\acmPrice{15.00}
\acmDOI{10.1145/3307630.3342384}
\acmISBN{978-1-4503-6668-7/19/09}

\usepackage[all=normal, paragraphs=tight]{savetrees}
\begin{document}
\title[MetricHaven --- More Than 23,000 Metrics for Measuring Quality Attributes of SPLs]{MetricHaven --- More Than 23,000 Metrics for Measuring\\Quality Attributes of Software Product Lines}

\author{Sascha El-Sharkawy}
\affiliation{%
  \institution{University of Hildesheim, Institute of Computer Science}
  \streetaddress{Universit{\"a}tsplatz 1}
  \city{Hildesheim}
	\country{Germany}
  \postcode{31134}
}
\email{elscha@sse.uni-hildesheim.de}

\author{Adam Krafczyk}
\affiliation{%
  \institution{University of Hildesheim, Institute of Computer Science}
  \streetaddress{Universit{\"a}tsplatz 1}
  \city{Hildesheim}
	\country{Germany}
  \postcode{31134}
}
\email{adam@sse.uni-hildesheim.de}

\author{Klaus Schmid}
\affiliation{%
  \institution{University of Hildesheim, Institute of Computer Science}
  \streetaddress{Universit{\"a}tsplatz 1}
  \city{Hildesheim}
	\country{Germany}
  \postcode{31134}
}
\email{schmid@sse.uni-hildesheim.de}

\begin{abstract}
Variability-aware metrics are designed to measure qualitative aspects of software product lines.
As we identified in a prior SLR \cite{El-SharkawyYamagishi-EichlerSchmid19}, there exist already many metrics that address code or variability separately, while the combination of both has been less researched.
MetricHaven fills this gap, as it extensively supports combining information from code files and variability models. Further, we also enable the combination of well established single system metrics with novel variability-aware metrics, going beyond existing variability-aware metrics.
Our tool supports most prominent single system and variability-aware code metrics. We provide configuration support for already implemented metrics, resulting in 23,342 metric variations. Further, we present an abstract syntax tree developed for MetricHaven, that allows the realization of additional code metrics.\\
\hspace*{-7pt}
\begin{tabular}{ll}
	\textbf{Tool:}&\url{https://github.com/KernelHaven/MetricHaven}\\
	\textbf{Video:}&\url{https://youtu.be/vPEmD5Sr6gM}
\end{tabular}
\end{abstract}

%
%


\maketitle

\section{Introduction}
\label{sec:Introduction}
In software engineering, metrics are an established approach to characterize properties of software \cite{FentonBieman14}. However, variability management is a key part of Software Product Lines (SPLs), which is not covered by traditional metrics. The SPL research community developed variability-aware metrics to address this issue, which received increasing attention over the last decade \cite{El-SharkawyKrafczykSchmid17}. While a few very specialized metrics have been integrated into SPL-specific IDEs, there is still a lack of publicly available metric tool suites. Here, we introduce MetricHaven, an extension of KernelHaven \cite{KroeherEl-SharkawySchmid18b}, for the flexible execution and combination of variability-aware code metrics.\trim

MetricHaven provides support for the measurement of traditional code metrics that ignore variability, SPL metrics that measure only variability, and the combination of both. Further, we allow the flexible combination of variability-aware code metrics with feature metrics, e.g., to detect complexity that arises through the combination of the variability model and code artifacts. MetricHaven provides configuration support to select among 23,342 metric variations (as of spring 2019). This is achieved by a specialized Abstract Syntax Tree (AST), which covers relevant information for measuring this broad variety of metrics. Through the underlying architecture of KernelHaven, which is optimized for parallelization, MetricHaven is able to measure a small number of selected code metrics on the whole Linux Kernel in less than 5 minutes, while the computation of all 23,342 metric variations takes around 3,5 hours. This means that in average each metric requires less than 1 second to measure the complete source code of Linux.

We regard MetricHaven as an experimentation workbench \cite{KroeherEl-SharkawySchmid18a} for the analysis of SPL metrics. The provided concepts support researchers in developing and evaluating new variability-aware metrics.
More precisely, we plan to analyze the potential of the metrics for fault prediction, by applying machine learning on the produced data.
Also, practitioners may benefit from MetricHaven as they can directly use the tool for the detection of code smells.

\section{Concept}
\label{sec:concept}
MetricHaven supports metrics for single systems, variability-aware metrics for SPLs, and arbitrary combinations of both. Further, we allow the combination of feature metrics for variability-aware metrics, if they count the number of features as part of their computation. As a result, MetricHaven supports more than 23,000 metric variations. Below we list the most important design decisions of MetricHaven:

\textbf{Decoupling of Parsers and Metrics.} The underlying architecture of KernelHaven provides data models, which serve as input for different kind of analyses \cite{KroeherEl-SharkawySchmid18b}. This allows reuse of extractors and data models for the realization of new analyses. For the implementation of MetricHaven, we could benefit from existing extraction capabilities for variability and build models. The extraction of the variability model and its data model required minor adaptations for storing hierarchy information, which serve as input for some of the feature metrics. Only new extractors are required for measuring product lines that use modeling techniques that are not supported so far.\trim

\begin{figure}[tb]
	\centering
		\includegraphics[width=\columnwidth]{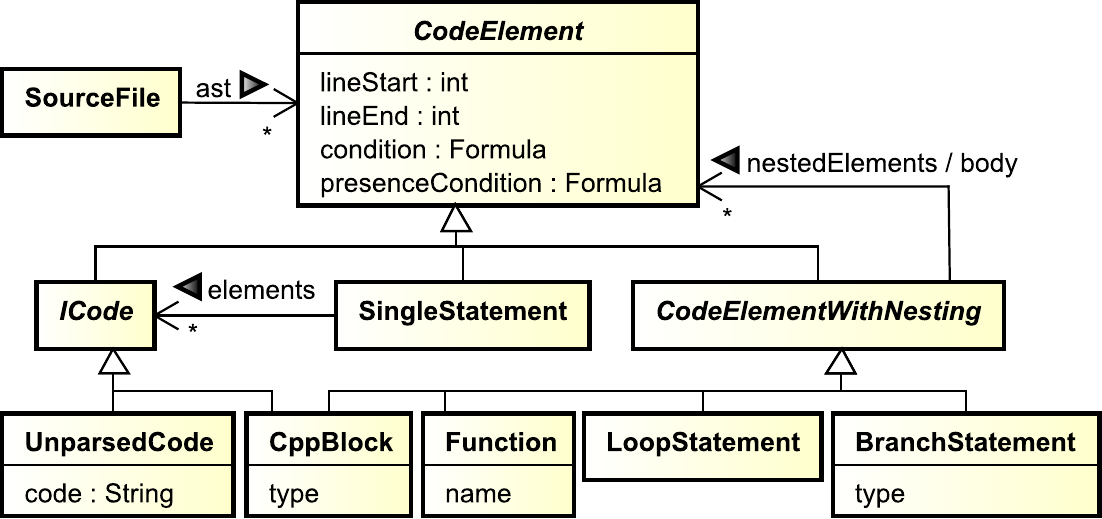}
	\vspace*{-15pt}
	\Description[Class diagram of MetricHaven's AST]{Class diagram of MetricHaven's abstract syntax tree (AST). A SourceFile contains abstract CodeElements, which are further refined. The abstract class CodeElementWith nested inherits from CodeElement and is the parent of CppBlock, Function, LoopStatement, and BranchStatement. Also the SingleStatement inherits from CodeElement and uses ICode elements to store unparsed elements of expressions. The already mentioned CppBlock also inherits from ICode element.}
	\caption{Excerpt of MetricHaven's AST.}
	\label{fig:ClassDiagram}
\end{figure}

\textbf{Common AST.} To allow the arbitrary implementation of single system and variability-aware code metrics, we require one common AST that stores elements of the annotation language, e.g., C-preprocessor, and elements of the programming language, e.g., C statements\footnote{The concepts we propose here, could also be applied well beyond C.}. This is a challenging task, since the preprocessor is not part of the programming language and can be used at arbitrary positions inside a code file, independently of any syntax definitions. On the other side, we do not need a syntactical correct and complete AST, since the AST should not be used for compilation tasks or type checking analyses. Figure~\ref{fig:ClassDiagram} presents an excerpt of our AST, which was motivated by our systematic literature study on variability-aware metrics \cite{El-SharkawyYamagishi-EichlerSchmid19} and an informal survey on traditional code metrics. The two most import elements are:
\begin{itemize}
	\item The \texttt{CppBlock} is used to store preprocessor blocks (\texttt{\#if}, \texttt{\#ifdef}, \texttt{\#ifndef}, \texttt{\#elif}, \texttt{\#else}). Since it inherits from two classes (\texttt{ICode} and \texttt{CodeElementWithNesting}), it allows the insertion of preprocessor directives at arbitrary positions in our AST.
	\item The \texttt{SingleStatement} is the most fine-grained element of our AST, which is sufficient for the measurement of the most prominent metrics. Expressions are stored as unparsed elements.
\end{itemize}

\textbf{Preprocessing Framework.}
The metric calculation operates on a per-function basis. This simplifies the implementation of metrics and also creates parallelization opportunities. However, some metrics (e.g.\ the Fan-In/-Out and Scattering Degree metrics) require a complete view on the whole code. For this, we implemented pre-processing components that run before the metric computation on the full models supplied by the extractors. The results of these pre-processing components are passed to the metric calculation component, which can look up these pre-calculated values. 

\subsection{Variability-aware Code Metrics}
\label{sec:Concept:Code Metrics}
The AST given in Figure~\ref{fig:ClassDiagram} stores enough information to realize traditional code metrics for single systems, variability-aware metrics for SPLs, and arbitrary combinations of both. Below, we present currently realized metrics, which may be selected by users to measure properties of SPLs (cf.\ Section~\ref{sec:Usage}). However, these metrics are not complete as the AST supports the definition of further metrics.

\textbf{McCabe's Cyclomatic Complexity Measure} \cite{McCabe76} is a very common metric for single systems that computes the number of linear independent paths of the control flow representation. This may be computed by adding 1 to the sum of all branching statements (\texttt{if}, \texttt{while}, \texttt{for}, \texttt{case}, \ldots) \cite{ConteDunsmoreShen86}. This approach was also applied to variation points to measure complexity of variation points \cite{Lopez-HerrejonTrujillo08}.
We support three alternatives to compute the cyclomatic complexity: The first counts only branching statements of the programming language. The second counts only variation points, e.g., \ifDefBlock{s}. 
The last one counts both kinds of branching statements, independently whether branching statements of the programming language are repeated in multiple variation points.\trim

\textbf{Nesting Depth (ND)} measures the maximum / average nesting level of statements within control structures \cite{ConteDunsmoreShen86}. The authors argue that each nesting increases the complexity as a developer needs to consider all enclosing conditions when editing the code. This concept was also applied to variation points \cite{LiebigApelLengauer+10}. We support three variations of this metric by counting the nesting depth of statements of the programming language, variation points, and the combination of both.

\textbf{Fan-In/-Out} metrics are designed to trace how data is processed inside a program \cite{ConteDunsmoreShen86}. Ferreira et al.\ \cite{FerreiraMalikKaestner+16} designed metrics that measure the incoming and outgoing connections of functions in combination with the number of features used for the selection of these connections. Based on these metrics, we defined 3 variations of \textit{Fan-In/-Out} metrics. The first variation measures only connections between functions and ignores the variability of the code. The second variation counts only conditional connections, that are function calls surrounded by at least one variation point inside the function. The third variation counts the number of features used to control the conditional inclusion of a function call.

\textbf{Lines of Code (LoC)} measures are very important metrics for single systems, but also in the context of SPLs. There exist a plethora of interpretations how to measure LoC. Jones \cite{Jones86} recommended to count statements instead of lines in order to have a formatting independent measure. \textit{Lines of Feature code (LoF)} measures the lines of code that are controlled by a variation point \cite{LiebigApelLengauer+10}. In addition, the \textit{Fraction of Annotated Lines of Code (PLoF)} computes $\frac{\text{LoF}}{\text{LoC}}$. We support LoC, LoF, and PLoF on statement level. 
Please note that the data model given in Figure~\ref{fig:ClassDiagram} also conceptually supports to measure the lines of code instead of statements, as we also store  the start and end line of each element.

\textbf{Features per Function} are designed to measure the complexity induced by the presence of \ifDefBlock{s} in the code \cite{FerreiraMalikKaestner+16}. We implemented 5 variations: 1.) Measures the number of features used inside a function. 2.) Measures the number of features that control the presence of the function, \textit{excluding} the build model. 3.) Measures the number of features that control the presence of the function, \textit{including} the build model. 4.) The union of the first and second metric. 5.) The union of the first and third metric.

\textbf{Blocks per Function} measure the number of variation points, instead of the features, used inside a code function \cite{FerreiraMalikKaestner+16}. The authors do not specify whether \texttt{\#else}/\texttt{\#elif}s are treated as separate blocks or as part of the \ifDefBlock. For this reason, we implemented two variations of this metric: The first metric counts only the number of \ifDefBlock{s} and ignores its siblings. The second treats related \texttt{\#else}/\texttt{\#elif}s as separate blocks.

\textbf{Tangling Degree (TD)} counts the number of features used in variation point expressions \cite{LiebigApelLengauer+10}. We realized only one version, as we do not know about a relevant variation of it.

\subsection{Feature Metrics}
\label{sec:Concept:Weights}
MetricHaven allows a flexible combination of code metrics with feature metrics. This can be done since most of the variability-aware code metrics count the number of features used to control the variation points. For these metrics we allow the usage of a feature metric to measure the complexity of the variation point instead of counting the number of used features only. Below, we introduce 7 feature metrics. Most of them are based on existing measures, but we developed also some new metrics for unmeasured properties.\trim

\textbf{Scattering Degree (\SD)} metrics are used to analyze whether a feature is implemented in a modularized way (low scattering) or is spread over the code base (high scattering) \cite{PassosQueirozMukelabai+18}. A high scattering indicates a fragile design and may significantly increase system maintenance. We re-implemnted two variations: \SDvp \cite{LiebigApelLengauer+10} is a commonly used metric that counts the number of variations points (i.e., \ifDefBlock{s}) a feature is used in. \SDfile \cite{HunsenZhangSiegmund+16} is a more coarse-grained version as it counts the number of code files, in which a feature is used in at least one variation point.

\textbf{Feature Size} is inspired by the LoF-metric of the previous section. We count for each feature the statements controlled by that feature. We count a statement to all features that \mbox{(in-)}directly control the inclusion of the statement. For instance, the statements in Lines~\ref{lst:code:if_start}--\ref{lst:code:if_end} of Figure~\ref{fig:Concept:ControlFlow} are controlled by the feature \textit{A}. The idea of this metric is to provide an alternative to the previously defined scattering degree metric, which shows the impact of a feature to the implementation of the SPL.

The \textbf{Number of Constraints (\NoF)} metric \cite{BagheriGasevic11} is often used to measure the complexity of feature models, by counting the number of cross-tree constraints. We refined this metric and provide 3 feature metric derivations: \NoF{} counts the number of all constraints of the feature model, where the respective feature is used in. 
Some modeling approaches allow a differentiation among constraints, e.g., the modeling approach of Linux requires that constraints are modeled as attributes of features \cite{El-SharkawyKrafczykSchmid15}. We allow a separate analysis of constraints modeled as attributes of a feature (\NoFOut) and constraints of other features referring to the measured feature (\NoFIn).

The \textbf{Coefficient of Connectivity-Density (\CoC)} is designed to measure how well graph elements are connected \cite{BagheriGasevic11}. This is done by dividing the number of edges (parent child relations and constraints) by the number of features in a feature model. We modified this metric to operate on features instead of the complete feature model only and provide two alternative measures: The first alternative counts the number of non-transitive child features, to measure the number of parent-child edges. We do not count the parents, since it would increase all numbers by one only, excerpt for the root feature. The second alternative considers also all incoming and outgoing constraint connections, which are also measured by the \NoF-metric of the previous Section.

The \textbf{Feature Types} metric is inspired by work that analyzes the amount of non-Boolean features used in publicly available SPLs \cite{PassosNovakovicXiong+11}. We allow to specify weights for each feature type. This facilitates an analysis whether the usage of non-Boolean features in variation points increases the code complexity.

\textbf{Feature Hierarchies} is inspired by work on how structural properties impact the complexity of feature models. This includes the measurement of the depth of tree, the number of top features, or the number of leaf features \cite{BagheriGasevic11}. We support two variations of hierarchy-based feature metrics: The first variation uses the nesting level of the feature as a feature metric. The second variation allows user-defined weights for top-level, intermediate, and leaf features.

This \textbf{Feature Locality} metric 
is a novel metric designed for the modeling behavior of Linux. Its variability model is spread over several hundred files \cite{BergerSheLotufo+13}. The used include mechanism provides a modularization concept at which a sub feature-model can be stored at the same location as the realizing implementation. In addition, it is possible to provide alternative sub feature-models containing the same features with different constraints for different CPU architectures. We provide a feature metric to measure the distance of a feature realization and the feature definition. This is done by computing the distance in the file system of the sub feature model that defines the feature and the measured code artifact, which uses the respective feature. If the feature is defined by multiple alternative files, we count the shortest distance.

\subsection{Example}
\label{sec:Concept:Example}
Based on the variable control flow representation shown in Figure~\ref{fig:Concept:ControlFlow}, we demonstrate how MetricHaven supports the computation of McCabe's cyclomatic complexity metric \cite{McCabe76} as well as the variability-aware version of the metric \cite{Lopez-HerrejonTrujillo08}. Further, we demonstrate how to combine the variability-aware code metrics presented in Section~\ref{sec:Concept:Code Metrics} with feature metrics presented in Section~\ref{sec:Concept:Weights}.

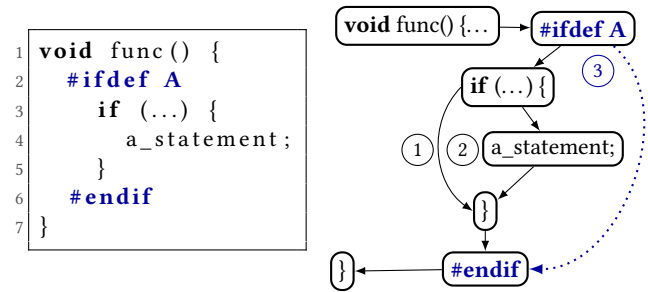
\begin{figure}
	\hspace*{-2ex}\begin{minipage}{.46\columnwidth}
\begin{lstlisting}[language=cWithPre]
void func() {
  #ifdef A%*\tRef{featureA}*)
    if (%*$\ldots$*)) { %*\label{lst:code:if_start}*)
      a_statement;%*\label{lst:code:call}*)
    }  %*\label{lst:code:if_end}*)
  %*\tRef{box2end}*)#endif%*\label{lst:code:block2end}*)
}
\end{lstlisting}
	\end{minipage}\hspace{10pt}
	\begin{minipage}{.51\columnwidth}
		\begin{tikzpicture}[start chain = going below, box/.style = {draw,rounded corners,fill=white, on chain,align=left, thick}]
			\node[box] at (0cm, 0cm)     (b1) {\textbf{void} func() \{\ldots};      
			\node[box] at (2.2cm, 1.2cm) (b2) {\bfseries\color{blue!60!black}\#ifdef A};    
			\node[box] at (1.2cm, .45cm) (b3) {\textbf{if} (\ldots) \{};										
			\node[box] at (1.8cm, -.4cm) (b4) {a\_statement;};														  
			\node[box] at (.9cm, -1.2cm) (b5) {\}};																					
			\node[box] at (.9cm, -2cm)   (b6) {\bfseries\color{blue!60!black}\#endif};			
			\node[box] at (-1cm, -2cm)   (b7) {\}};     
		 
			\node[draw, circle, inner sep = 2pt] at (0cm, -1.65cm) (path1) {\small 1};
			\node[draw, circle, inner sep = 2pt] at (.6cm, -1.65cm) (path2) {\small 2};
			\node[draw=blue!60!black, circle, inner sep = 2pt] at (2.4cm, -.6cm) (path3) {\small \color{blue!60!black}3};
		
		 \begin{scope}[rounded corners,-latex]
			 \path (b1) edge (b2) (b2) edge (b3) (b3) edge (b4) (b4) edge (b5) (b5) edge (b6) (b6) edge (b7);
			 \path (b2.-30) edge[dotted, bend left=70, draw=blue!60!black, thick] (b6.0);
			 \path (b3.-180) edge[bend right=50] (b5.180);
			\end{scope}
		\end{tikzpicture}
	\end{minipage}
	\vspace*{-5pt}
	\Description[Example of a C- code function and its variable control-flow representation, consisting of seven nodes and 3 linear independent paths.]{Example of a C-code function, containing one if statement surrounded by a C-preprocessor block. One statement is nested inside both structures. The right side shows the variable control-flow representation, consisting of seven nodes and 3 linear independent paths.}
	\caption{Code example and its variable control-flow.}
	\label{fig:Concept:ControlFlow}
\end{figure}

\textbf{McCabe on Code.} McCabe's metric \cite{McCabe76} computes the linear independent paths of the control flow representation of a program. Our example contains two linear independent paths, drawn with black lined arrows. This can be counted by adding 1 to the number of \texttt{Branch}- and \texttt{LoopStatement}s.

\textbf{McCabe on Variation Points.} The variability-aware version of McCabe considers only paths created by variation points \cite{Lopez-HerrejonTrujillo08}. The example contains two paths. The black lined arrows are treated as one path (keeping the \ifDefBlock) and the blue dotted arrows (removal of the \ifDefBlock). This can be counted by adding 1 to the number of \texttt{CppBlock}s.

\textbf{McCabe on Variation Points and Code.} Further, MetricHaven allows a combination of the single system and the variability-aware code metric in a manner, which was not done before. For the combination of both metrics, we need to count all three paths of the control flow graph.

\textbf{Combination with Feature Metrics.} McCabe proposed to treat compound expressions in the form of \texttt{Exp$_1$ and/or Exp$_2$} as separate paths as they could alternatively be realized with multiple if-then-else blocks \cite{McCabe76}. Following his idea, we support measuring of features in conditions rather than counting variation points only. Instead of adding 1 for each measured feature, we add the complexity value of a selected feature metric of Section~\ref{sec:Concept:Weights}. For example, we facilitate using the number of modeled constraints (\NoF-metric) in which feature A appears in instead of counting the blue dotted path.

\vspace{-3pt}
\section{Usage}
\label{sec:Usage}

MetricHaven is implemented as a plug-in for the KernelHaven infrastructure, which requires the Java runtime version 8 or higher. We recommend at least 16 GiB of RAM for the computation of all 23,342 metric variants. The default release of KernelHaven includes MetricHaven and required extractors for analyzing the Linux Kernel up to version 4.17. The variability and build model extractors work only on a Linux system with an installed GCC compiler plus the prerequisites for compiling the Linux Kernel. Other extractors may be supplied as additional plug-ins that allow analyzing other SPLs, which may not have these limitations.

An execution of MetricHaven requires a KernelHaven release\footnote{\url{https://github.com/KernelHaven/KernelHaven}} to be downloaded and unpacked.
The release includes the \texttt{code\_metrics.properties} configuration file, which can be used as a basis for the execution of MetricHaven.
The KernelHaven process is started with the following command\footnote{More details in appendix of \url{https://sse.uni-hildesheim.de/en/research/publications/publikation-einzelansicht/?lsfid=41623&cHash=b244065d45b8c85e2cd7560559dda3dd}}:
\vspace{-4pt}
\Command{\small\texttt{java~-jar~KernelHaven.jar~code\_metrics.properties}}
\vspace{-6pt}

KernelHaven's pipeline and the analysis components may be configured via Java properties files.
The most relevant settings with respect to MetricHaven are:
\begin{itemize}
	\item \texttt{source\_tree} specifies the path to the SPL to analyze. For example, this can be the root directory of an unpacked Linux Kernel.
	\item By default, all 23,342 metric variations are executed. \texttt{me\-trics\-.co\-de\-\_metrics} may be used to select all variations of one metric. The provided configuration file contains a list of supported metrics. Further, \texttt{metrics.fun\-ction\_mea\-sures.all\-\_vari\-a\-tions} may be used to select a single metric variation.\trim
	\item \texttt{code.extractor.threads} controls the number of threads used for parsing the source code of the product line.
	\item \texttt{metrics.max\_parallel\_threads} controls the number of threads used for calculating the metrics for a single function.
	This may be used for the computation of a large number of metric variations, otherwise the overhead of multithreading is too high.\trim
\end{itemize}

\section{Conclusion}
\label{sec:Conclusion}
We presented MetricHaven, an extension of KernelHaven for measuring metrics for single systems, variability-aware metrics for SPLs, and arbitrary combinations of both. Further, it supports the combination of feature metrics with variability-aware code metrics. Its flexibility allows the combination of 23,342 metric variations. This is achieved by an common AST that combines information of parsed code annotations (e.g., C-preprocessor) and the programming language (e.g., C-code).

MetricHaven is very fast. For example, it requires 3,5 hours for measuring all 23,342 metric variations on the complete Linux kernel\footnote{Measurement of Linux version 4.15 on 2 Xeon E5-2650 v3 with 128 GB memory}. This is about half a second per metric. Or, to put it differently, measuring 23,342 metrics, requires about 40ms per code function.

MetricHaven is a publicly available experimentation workbench designed to support researchers and practitioners in the analysis of annotation-based SPLs. New metrics may be developed based on the presented AST, without the need to develop new parsers. On the other side, the existing metrics may be applied without any changes to other imperative code files through the implementation of new extractors. This supports researchers in evaluating new SPL metrics. Also practitioners may benefit from MetricHaven as they can directly use the tool for the detection of code smells of annotation-based SPLs.

\begin{acks}
This work is partially supported by the ITEA3 project $\text{REVaMP}^2$, funded by the \grantsponsor{01IS16042H}{BMBF (German Ministry of Research and Education)}{https://www.bmbf.de/} under grant \grantnum{01IS16042H}{01IS16042H}. Any opinions expressed herein are solely by the authors and not of the BMBF.
\end{acks}

\bibliographystyle{ACM-Reference-Format}
\bibliography{literature}

\end{document}